# A Cross-Domain TARA-to-NAC Rule Mapping Methodology: From Automotive OEM Networks to Energy Storage Systems

Krishna Teja Medam, Department of Electrical Engineering & Computer Engineering, and Computer Science, University of Detroit Mercy, Detroit, MI, USA. kmedam@udmercy.edu

Thiago Quinelato, Department of Electrical Engineering & Computer Engineering, and Computer Science, University of Detroit Mercy, Detroit, MI, USA. tquinel@udmercy.edu

## Abstract

Threat analysis and risk assessment, as required by ISO/SAE 21434, identifies cybersecurity goals for automotive systems, but existing frameworks offer little structured guidance for translating those goals into enforceable network access control rules. Compliance standards such as IEC 62443, ISO/SAE 21434, and ISO/IEC 27001 are typically applied as parallel, loosely coordinated efforts rather than a unified enforcement architecture. This paper presents a systematic six-step methodology for mapping cybersecurity goals derived from a threat analysis and risk assessment to concrete network access control rules, each placed at a named enforcement point and traceable to the threat records behind it. It also presents a unified compliance framework that treats IEC 62443 and ISO/IEC 27001 as a fixed substrate and domain-specific regulation as an interchangeable overlay. The methodology is applied to a primary worked example of an automotive original equipment manufacturer plant network, addressing multi-tier supplier authorization and legacy operational technology device constraints. Its cross-domain applicability is then demonstrated through a secondary worked example in utility-scale battery energy storage, in which the North American Electric Reliability Corporation Critical Infrastructure Protection standards and ISO/IEC 27019 replace the automotive-specific standard while the substrate is held constant. In the automotive example, 37 threat entries and 24 cybersecurity goals yield 23 network access control rules covering 11 of the 13 in-scope goals; the energy example yields six domain rules. Three of the derived rules have no equivalent in the cited standards. Both worked examples are desk-based, and the methodology is applied to them rather than tested on them. The result is a reusable pattern for translating risk assessment into enforcement across critical infrastructure sectors. Broader applicability is discussed, and directions for empirical validation are outlined as future work.

## Introduction

A threat analysis and risk assessment (TARA) under ISO/SAE 21434 ends with a set of cybersecurity goals [1]. The goals say what must be protected and how much risk must be reduced. They do not say which rule, on which network device, will do the protecting, and the standard does not ask anyone to work that out in a traceable way [1]. Published TARA methods stop at the same place, with goals, requirements, or design patterns [2-5]. The one line of work that does derive rules from TARA artifacts does so for the in-vehicle CAN network, with every rule landing at a single gateway [6].

Compliance work tends to widen this gap rather than close it. An automotive OEM answers to ISO/SAE 21434 for its TARA [1], to IEC 62443 for its industrial zones and conduits [7,8], and to ISO/IEC 27001 for its management system [9]. Each is usually handled in its own effort and produces its own documents. Published mappings between these standards yield control catalogs or pairwise comparisons [10-13], and none follows a control down to a rule that a specific device enforces. What gets lost is the line from a deployed rule back to the risk that justified it.

The problem is sharpest where operational technology meets outside parties. A plant gives suppliers in several tiers remote access, and some of the devices those suppliers can reach are legacy PLCs that accept Modbus TCP commands from any sender. An identity check at the boundary decides who gets in, but once a session is in, the same check cannot tell a register read from a register write. This is a present concern. A joint advisory from NSA, CISA, FBI, DOE, and EPA (AA26-231A, Aug. 2026) reports reconnaissance and AI-assisted exploitation-tool development aimed at internet-exposed Siemens S7 Series PLCs, names Critical Manufacturing and Energy among the most-targeted sectors, and notes that PLC targeting reaches beyond a single vendor [14].

This paper makes three contributions. First, it presents a systematic methodology for mapping TARA-derived cybersecurity goals to concrete NAC enforcement rules.

Second, it presents a unified compliance framework that treats IEC 62443 and ISO/IEC 27001 as a fixed substrate with domain-specific regulation as an interchangeable overlay. Third, it applies the methodology to a primary automotive OEM worked example, addressing multi-tier supplier authorization and legacy OT device constraints, and demonstrates its cross-domain applicability through a secondary worked example in utility-scale battery energy storage, substituting NERC CIP and ISO/IEC 27019 for the automotive-specific standard while holding the substrate constant. Both worked examples are desk-based. The method is applied to them rather than tested on them. A companion paper evaluates an implementation of the automotive enforcement points, and none of its results are used here.

The rest of the paper is organized as follows. The Related Work section reviews prior work, and the Background section summarizes the parts of ISO/SAE 21434, IEC 62443, and ISO/IEC 27001 that the method relies on, along with the enforcement primitives and the energy overlay standards. The Methodology section sets out the mapping procedure, the rule taxonomy, and the test for rules that have no equivalent in the standards. The Unified Standards Compliance Mapping section presents the substrate/overlay framing, its crosswalks for the automotive overlay and the energy overlay, and the outcome for the three flagged rules. The two Worked Example sections work through the automotive and energy examples, and the appendix gives the full automotive TARA (Tables A1 to A4) and a comparison with prior work (Table A5). The Discussion section discusses what carried over between them, the Limitations section lists the limitations, and the final section concludes.

## Related Work

This section organizes prior work into seven strands, asking of each how far it carries a risk assessment toward something a network actually enforces.

### TARA Methods and ISO/SAE 21434

Automotive TARA is a mature discipline. Ward et al. carried hazard-analysis practice into cybersecurity, and the impact and feasibility ratings they used anticipated the structure later adopted by ISO/SAE 21434 [3]. Luo et al. provide the standard survey of method families [4], and Wang et al. show how cybersecurity requirements can be read directly off a risk matrix [5]. Dobaj et al. take the process one step further, carrying TARA results into design patterns for networked embedded systems [2]. Automation has arrived at the front end of the process: ThreatSurf generates and rates attack paths in an ISO/SAE DIS 21434-compliant way, which feeds the start of a TARA but not its end [15]. Practice does not paint a rosier picture. In interviews with 15 practitioners, Jing et al. found the TARA process labor-intensive for want of automated tools and judged the available threat catalogs unsatisfactory [16].

Wagner aggregates TARA output across divisions as risk metadata [17], which is useful for governance but produces no enforcement rule either. Across all of these, the TARA terminates at goals, requirements, or design patterns, and turning a goal into a configured rule at a specific network location is left to whoever implements the system. That unclaimed step is where this paper operates.

### From Risk Analysis to Enforced Rules

The nearest prior work is Navarra et al. [6], who derive CAN firewall rules from TARA artifacts through an automated, LLM-assisted workflow with explicit risk traceability. Their contribution is real, but it is bounded: every rule lands at the same enforcement layer (the vehicle gateway), the rule model has no place for suppliers or any other organizational boundary, and no compliance mapping beyond ISO/SAE 21434 is produced. Earlier, Luo et al. designed gateway security mechanisms from threat analysis but without documenting a derivation procedure [18,19], and Kleberger et al. partitioned in-vehicle networks into security domains algorithmically, stopping at the architecture [20]. Djebbar and Sogbade map threats to controls drawn from ISO/IEC 27001, NIST CSF, and IEC 62443-3-3, but what comes out is a list of controls to implement, not configured rules, and no TARA feeds the process [12].

Four things separate this paper from [6]. A single goal here can yield rules at two enforcement points rather than one. The rules govern tiered supplier identities crossing an organizational boundary, which the in-vehicle rule model does not represent. Every rule is embedded in a substrate/overlay compliance mapping that [6] does not produce. And two of the derived rule types, policy-version binding and deny-by-default for unclassified roles, have no equivalent in the cited standards. This paper also states plainly what it does not do: derivation here is manual, the automation demonstrated in [6] is future work, and the BESS example is a cross-domain transfer [6] never attempts.

### IEC 62443 Zones, Conduits, and Security-by-Design

IEC 62443 frames industrial security through zones, conduits, and security levels. Jaatun et al. observe that guidance on implementing conduits is thin and propose both principles and a process for connecting zones whose security levels differ [21]. Elmarkez et al. segment a system into zones and conduits on the basis of a threat evaluation, then identify security functions per zone [22]. Model-driven research formalizes the zone model itself, once as a domain-specific language with validation rules [23] and once as data-flow-driven zone modeling during system design [24]. What none of these produces is the set of per-identity rules a conduit actually enforces. Security functions are identified at the level of the zone, not traced down to individual rules at a named enforcement point.

The method in this paper supplies that missing traceable path: each rule is bound to an IEC 62443-3-3 requirement and to the ISO/SAE 21434 goal that justified it.

### Access Control in Industrial Networks

Access control policy for industrial networks has been studied thoroughly, but almost always from the policy downward. Cheminod et al. formalize how such policies are described [25] and verify semi-automatically that an implementation conforms to its specification [26]. Leander et al. compare enforcement architectures suited to dynamic manufacturing [27]. Kern and Anderl enforce attribute-based policy at both network and machine level [28], and P4Ward enforces device-specific behavioral policies at line rate [29]. In every case the policy is assumed to exist. Where it comes from, and how each individual rule earns its place against a risk assessment, remains an open question. That upstream step is what this paper contributes, and the consistency checks of [26] are a natural model for checking whatever rule set it generates.

### Multi-Standard Mapping and IT-OT Convergence

Standard-to-standard mappings are common. Costantino et al. compare UNECE R155 with ISO/SAE 21434 [10]. Akkus and Annighoefer compare aviation and automotive standards [13]. PURITY maps controls from ISO/IEC 27001, IEC 62443, NIST SP 800-82, and ISO/IEC 27033 onto Purdue-model levels, but connects none of that to a domain risk assessment and does not cover ISO/SAE 21434 [11]. Djebbar and Sogbade correlate threats to controls across ISO/IEC 27001:2022, IEC 62443-3-3, and NIST CSF [12], though the result stays at the level of a control catalog and nothing in it is anchored to a risk-derived rule at a named enforcement point. The mapping proposed here is anchored differently: to TARA goals and the rules derived from them, with IEC 62443 and ISO/IEC 27001 as the fixed substrate and domain regulation applied as an exchangeable overlay.

### Cross-Domain Transfer of Automotive Security Methods

Cross-domain transfer of automotive security methods is established practice, and the transfers have mostly run in one direction. Eschlberger et al. adapt a smart-grid security-analysis domain-specific language to the automotive domain under ISO/SAE 21434 and UN R155 [30], the reverse of the direction taken here. Rabe and Wollert find that automotive software-update standards transfer to fire alarm systems, and they name industrial automation under IEC 62443 as the next candidate domain [31]. Transfer alone, though, is not a contribution. What this paper transfers is a rule-derivation procedure together with its standards overlay, exercised end to end on the BESS case.

### Battery Energy Storage System Security

BESS security is well documented. Zografopoulos et al. survey threat modeling for cyber-physical energy systems [32]. Trevizan et al. review the cyber-physical security of grid-scale BESS [33]. Öhrström et al. demonstrate the attack surface of cloud-controlled BESS and its consequences for grid stability [34]. Kharlamova et al. review detection and mitigation [35], and Rafy et al. study cyber-resilient IoT-based BESS controllers [36]. None of this work, however, carries a structured risk assessment through to access control rules, which is exactly the gap the second worked example addresses.

### Summary and Research Gap

Existing work offers mature TARA methods, in-vehicle rule generation, framework-level threat-to-control mapping, formal access control policy, and pairwise standard comparisons. Within the literature reviewed here, none offers a reproducible procedure that carries TARA cybersecurity goals through to concrete NAC rules and enforcement points across domains. Table A5 in the appendix summarizes the positioning.

## Background

### TARA under ISO/SAE 21434

ISO/SAE 21434:2021 defines cybersecurity engineering for road vehicles and assigns the organization a TARA for items in scope [1]. For this paper the relevant chain is the one the mapping consumes, not the full standard. The item is defined. Assets are identified. Threat scenarios against those assets are stated. Impact and attack feasibility are rated, a risk value is determined, and a risk treatment decision follows. Where the decision is to reduce the risk, the concept phase produces cybersecurity goals. Those goals are the input to the Methodology section.

Two limits of that chain matter here. The standard specifies the assessment and the work products. It does not specify the enforcement technology, and it does not require that a goal be traced to a rule at a named point in the network. Published method work confirms the stopping point. Impact and feasibility ratings give the assessment its structure [3]. Method surveys organize the families of analysis [4]. Requirements can be read off a risk matrix [5], and the results can be carried into design patterns for networked embedded systems [2]. Automation has reached attack-path generation at the start of a TARA [15]. Practitioner accounts describe the process as labor-intensive and the available threat catalogs as weak [16]. Cross-divisional practice aggregates the output as risk metadata [17]. None of these steps emits a NAC rule.

The mapping in the Methodology section therefore starts where this chain stops. A cybersecurity goal is accepted as given, together with the asset, the threat scenario, and the treatment decision that produced it. The worked examples use summarized TARA outputs; the full automotive analysis is in the appendix (Tables A1 to A4).

### IEC 62443 Zones, Conduits, and the Requirements a Rule Can Satisfy

IEC 62443 is the industrial automation and control standard against which the rules are bound. Two parts are in scope. IEC 62443-2-1:2024 states security-program requirements for the asset owner [8]. IEC 62443-3-3:2013 states system security requirements and security levels [7]. The security level vector, as used in recent 62443 implementations [21,23], expresses a target per foundational requirement, but security levels express the target a zone is designed to meet; they do not by themselves select a rule. The architectural objects used to allocate those requirements are the zone and the conduit. A zone groups assets that share security requirements. A conduit is the controlled path between zones, or between a zone and an external network. Jaatun et al. note that conduit implementation guidance is thin and propose principles for connecting zones of different security levels [21].

Other work segments a system from a threat evaluation and identifies security functions per zone [22], or formalizes the zone model during design [23,24]. Security functions identified per zone are not yet the per-identity rule set a conduit enforces. That distinction is the reason the Methodology section binds each derived rule to a 62443-3-3 requirement and to the goal that justified it.

The requirements used later are the ones a NAC rule can actually discharge. Identification and authentication control, in Foundational Requirement 1 (FR 1), is the identity-layer decision: the session presents a principal the conduit will recognize. Use control, FR 2, is the command decision: an authenticated principal is still limited in what it may invoke. Restricted data flow, FR 5, is the segmentation decision: the conduit carries only the flows the zone model allows. Security levels express the target a zone is designed to meet; they do not by themselves select a rule. Legacy devices that cannot hold a credential break FR 1 at the endpoint. The automotive example treats that case as a placement problem (enforcement moves to the conduit) rather than as an exception to the goal.

### ISO/IEC 27001 as the Management Substrate

ISO/IEC 27001:2022 specifies an information security management system (ISMS): scope, risk assessment, risk treatment, and a control set the organization must justify [9]. It is not an OT network architecture, and it does not define zones, conduits, or protocol functions. It is the substrate because the obligation it imposes does not change when the domain overlay changes. An automotive supplier program and a BESS operator both need a declared scope, a risk treatment decision, and evidence that the implemented control matches the decision. IEC 62443-3-3 supplies the system requirement the rule is written against [7]. ISO/IEC 27001 supplies the management obligation under which that rule is approved, reviewed, and retained as evidence [9].

This split is deliberate. Mappings keyed by control, with 27001, 62443, and a domain standard side by side, produce a control catalog [11,12]. The frame in the Unified Standards Compliance Mapping section is keyed by rule. A rule is first a discharge of a 62443-3-3 requirement, traced to a TARA goal. The 27001 obligation is the evidence and review duty attached to that rule. Domain regulation is applied only after both substrate bindings exist.

### Enforcement Primitives

The identity layer in the mapping is port-based network access control as operated with RADIUS under RFC 3580 [37]. A principal is authenticated at the point of attachment, and the authorization returned to the authenticator selects the segment or conduit the session may enter. That decision admits the session. It does not inspect the industrial function the session then invokes.

The protocol layer is the second decision. NIST SP 800-82 Revision 3 treats OT networks as systems in which availability and safety constrain the controls that can be applied, and in which segmentation and controlled remote access are primary defenses [38]. NIST SP 800-53 Revision 5 supplies the control language for access enforcement and least privilege against which an information-system rule can be stated [39]. Neither publication derives the rule from a TARA. Industrial access-control research assumes the policy already exists and then formalizes it [25], checks an implementation against it [26], or enforces it at the machine or at line rate [28,29]. The Methodology section is the upstream step those results leave open: the procedure that produces the policy, with one rule at the identity layer and, where the goal concerns actuation, a second rule on the protocol function.

The CISA advisory illustrates why the second rule is not redundant. Exposed S7 controllers were reached over S7comm, and the activity of concern included read and write operations on data blocks and program logic [14]. An identity check at the perimeter would not have distinguished a read used for monitoring from a write used to change the process. The mapping therefore treats the protocol function as a separate authorization object.

### Overlay Standards for the Energy Example

The Worked Example 2 section replaces the automotive overlay. It does not replace the substrate. Two instruments carry the energy overlay.

NERC CIP applies to bulk electric system (BES) cyber systems, and applicability follows categorization rather than the mere presence of storage [40,41]. An electronic security perimeter is not a 62443 conduit. Interactive remote access is not the same object as an on-site supplier session. A BESS that is not a categorized BES cyber system does not inherit the medium- or high-impact obligations. The example therefore states the assumed impact category before any CIP identifier is attached to a rule. CIP is used as the overlay obligation, not as a second copy of the 62443 binding.

ISO/IEC 27019 provides information security controls for the energy utility industry, based on ISO/IEC 27002:2022, for monitoring and controlling the production, transmission, storage, and distribution of electric power, and the supporting process-control systems [42]. Its scope includes digital controllers, PLCs, protection and safety systems, remote maintenance, and energy management for distributed resources and storage. It is sector guidance for the control set. It is not a TARA method and not a rule generator. In the overlay it specializes the 27001 evidence duty for utility process control; it does not replace FR 1, FR 2, or FR 5.

## Methodology: TARA-to-NAC Rule Mapping

The method takes the outputs of a TARA and returns NAC rules. Each rule sits at a named enforcement point and traces back to the threat that justified it. The method does not repeat the TARA. Its input is the set of cybersecurity goals and the threat records that produced them (see TARA under ISO/SAE 21434); its output is a rule table, which the Unified Standards Compliance Mapping section binds to the standards. Derivation is manual, and each step is written down so that a second analyst given the same TARA can reproduce the rule set, or point to exactly where their judgment departs from the derivation given here.

### Mapping Procedure

Two records carry the procedure. A threat record is $t = (f, a, s, d, i, g)$: function, asset, STRIDE category, damage scenario, impact rating, and the goal it produced. One goal can come from several threat records. A rule record is $r = (id, G, c, act, p, l, b)$: an identifier, the set of goals the rule serves, the trigger condition, the action, the enforcement point, the rule layer (see Rule Taxonomy), and the binding to a standard requirement, or a flag where none exists (see Rules Without a Standard Equivalent). Triggers are things an enforcement point can actually observe: an authentication result, a certificate role attribute, a protocol function code, a change to configuration. Actions are admit, assign, deny, quarantine, log, and hold.

The procedure has six steps (Figure 1, Algorithm 1).

*Step 1, select goals.* Every goal with a risk-reduction treatment enters the procedure. Goals deferred to future work stay in the goal table (Table A4) as unmapped; they are reported, not dropped.

*Step 2, ask three questions per goal.* What is protected: the asset and the damage scenario. Who is the subject: the principal, role, or session whose request reaches the asset. How can a network decision enforce the goal. If no network decision can discharge the goal, as with firmware integrity inside a device, the goal is marked out of NAC scope with a reason.

*Step 3, place the rule.* The enforcement point is chosen on the path from subject to asset. A goal about who may reach an asset is placed at the identity layer, at the point of attachment [37], or at the zone boundary. A goal about what an admitted subject may do to the asset, such as actuation or writes, also gets a rule at the protocol layer. Where the asset cannot hold a credential, placement moves to the conduit in front of it (see IEC 62443 Zones, Conduits, and the Requirements a Rule Can Satisfy). One goal can therefore produce rules at two enforcement points.

*Step 4, write the rules.* Each distinct trigger at an enforcement point gets one rule. Every trigger class gets a default branch for the no-match case, so an input the policy does not recognize is denied or quarantined rather than passed. Every decision rule gets an audit rule. Each rule is bound to the standard requirement it discharges [7,8,37-39]. A rule with no such requirement is flagged.

*Step 5, consolidate.* Rules from different goals with the same trigger, action, and enforcement point are merged. The merged rule keeps every goal it serves, so traceability becomes many-to-many. Where two rules share a trigger and enforcement point but differ in action, the more restrictive action is kept and the conflict is recorded against both goals.

*Step 6, check.* Three checks close the procedure. Backward trace: every rule names at least one goal and, through it, a threat record. Forward trace: every goal in scope has at least one rule, or is listed as unmapped with a reason. Default coverage: every trigger class has a deny or quarantine branch.

Checking the deployed configuration against the rule table is left to tools of the kind described in [26].

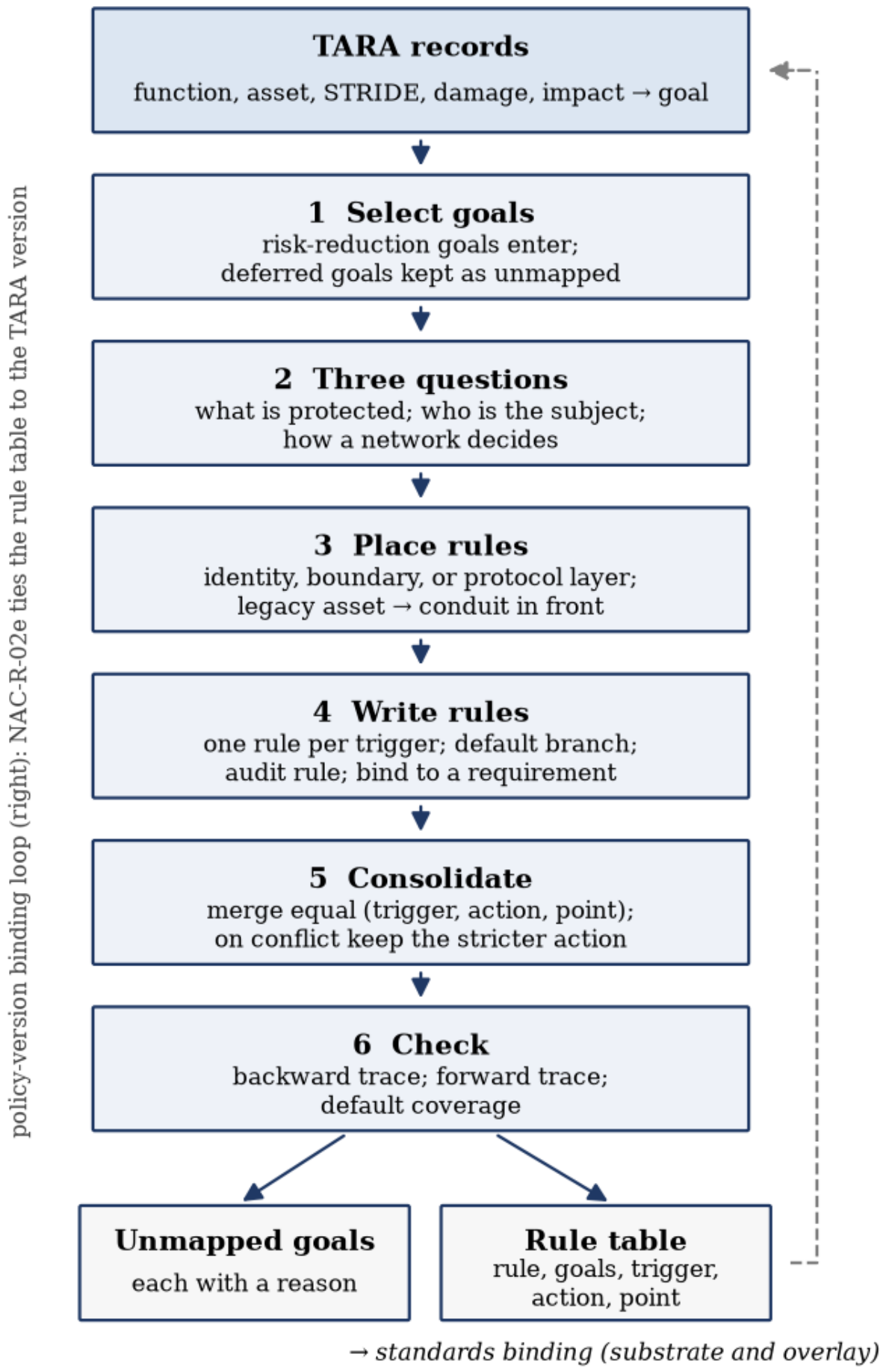

**Figure 1.** TARA-to-NAC mapping procedure. Each rule keeps a backward trace to its threat records; policy-version binding ties the deployed rule set to the TARA version.

```
Algorithm 1. TARA-to-NAC rule derivation
Input: goals G, threat records T
Output: rule table R, unmapped list U
R <- {}; U <- {}
for each g in G, treatment = reduce:
  (asset, damage) <- what(g, T)
  S <- who(g, T)
  P <- place(g, S)
  if P empty: add (g, reason) to U
  for each point p in P:
    for each trigger class c at p:
      add rule c -> action, per s in S
      add default c -> deny|quarantine
    add audit rule at p
    bind rule to requirement, else flag
merge equal (trigger, action, p)
on conflict keep stricter action
check back trace, fwd trace, defaults
return R, U
```

### Rule Taxonomy

The rules fall into two categories, each defined by the decision it makes rather than by the product that enforces it (Table 1).

Identity and segmentation rules decide whether a session is admitted and into which zone. They cover admission, role-based zone assignment, boundary enforcement between zones, and quarantine. They discharge FR 1, FR 2, and FR 5 of IEC 62443-3-3 [7] and are enforced at the RADIUS server and the zone boundary [37].

Protocol and command rules decide which industrial function an admitted session may invoke. They cover permitted reads, denied writes, and fail-closed handling of frames the enforcement point cannot parse. They discharge FR 2, use control [7], in the least-privilege sense of [39], and are enforced at a protocol-aware proxy in front of the device.

A supporting class protects the rule set itself: audit, change control, integrity checks, and policy-version binding. These rules make no session decision. They keep the other two categories trustworthy over time, and they carry the ISO/IEC 27001 evidence duty described in the ISO/IEC 27001 as the Management Substrate subsection [9].

The categories carry across domains because they name decisions, not protocols. In the Worked Example 2 section the protocol and the roles change; the two decision layers do not.

**Table 1.** Rule taxonomy.

| Category | Decision | Enforcement point | IEC 62443-3-3 |
|---|---|---|---|
| Identity and segmentation | Admit; assign zone by role; enforce zone boundary; quarantine | RADIUS server; zone boundary | FR 1; SR 2.1; SR 5.2 |
| Protocol and command | Permit read; deny write; fail closed on unparsed frames | OT protocol proxy | SR 2.1 (FR 2) |
| Supporting | Audit; change control; integrity check; version binding | Policy host; audit log | SR 2.8; SR 3.4; 62443-2-1 CM 1.4 |

### Rules Without a Standard Equivalent

A rule is flagged when no requirement in the cited standards mandates its trigger and action together, even where each part is supported somewhere. The flag does not claim the rule is novel practice. It records that a reader cannot reach the rule by searching the standards. Two automotive rules meet this test.

*NAC-R-02e, policy-version binding.* Each deployed NAC policy carries the version of the TARA goal set it was derived from. When the two differ, the mismatch is flagged and the policy is not put into force until it is reconciled with the current goals. Change control governs how a rule changes [8,39], but nothing ties the rule to the risk assessment that justified it. Without that tie, a TARA can be revised while the network still enforces the old goals.

*NAC-R-08d, deny-by-default for unclassified roles.* A principal that authenticates successfully but carries a role attribute matching no defined tier is placed in quarantine, and an alert is raised. The standards cover failed authentication and deny-by-default between zones [7,38]. They do not address a valid credential with a role the policy does not know. In a tiered supplier program that case arises whenever a tier is added, a role is renamed, or a certificate is issued from a template the policy was not written for.

Neither rule depends on the automotive domain. Policy-version binding applies to any rule set derived from a risk assessment. Deny-by-default for unclassified roles applies wherever external parties are tiered by role. In the energy example, it is a vendor credential whose role maps to neither the power-conversion nor the battery-management vendor zone, which NAC-R-E-06 sends to quarantine. The Worked Example 2 section reports whether any energy rule meets the test on its own.

## Unified Standards Compliance Mapping

### Substrate and Overlay

The Methodology section produces a rule and, where one exists, a binding to a requirement the rule discharges. This section decides that binding. Two layers are kept apart. The substrate does not change between the examples. The overlay does.

The substrate has two parts, and they are not the same kind of obligation. IEC 62443-3-3 supplies the system requirement a session or command rule is written against: identification and authentication (FR 1), use control (FR 2), and restricted data flow (FR 5) [7]. IEC 62443-2-1 supplies the asset-owner program duty for the supporting rules, in particular change control under CM 1.4 [8]. ISO/IEC 27001 supplies the management obligation that does not change when the domain changes: a declared scope, a risk-treatment decision, and evidence that the implemented control matches the decision [9]. A rule is first a discharge of a 62443 requirement. The 27001 control is the evidence and review duty attached to that rule. Treating 62443 and 27001 as one substrate is accurate only if that split is kept. A table keyed by control, with 62443, 27001, and a domain standard side by side, is the parallel catalog this paper is written against [11,12]. Tables 2 and 3 are keyed by rule.

The overlay is the domain obligation applied after both substrate bindings exist. In the Worked Example 1 section, it is ISO/SAE 21434 [1]. In the Worked Example 2 section it is NERC CIP-005, under the categorization of CIP-002, and ISO/IEC 27019 [40-42]. The overlay does not replace FR 1, FR 2, or FR 5, and it does not generate the rule. It states the domain duty the already-derived rule also serves.

A substrate binding names the requirement a rule discharges. An overlay entry names a duty the rule also serves. Neither has to state the rule's trigger and action. The flag test is stricter: a cited requirement must cover the trigger and the action together, and support for one half is noted without removing the flag. A rule with no such requirement keeps the flag of the Rules Without a Standard Equivalent subsection.

### Automotive Overlay

The automotive overlay is ISO/SAE 21434, applied here to a non-vehicle item, the plant network [1]. Clause 9 produces the cybersecurity goals the procedure consumes (WP-09-03). Clause 7 frames the supplier tiers through the cybersecurity interface agreement (WP-07-01), which is why a tier is a principal the rules can name. Neither clause specifies a NAC rule, an enforcement point, or a protocol function. The 27001 column is the evidence duty, not a second system requirement.

NAC-R-02e has no match. Trigger: the deployed NAC policy version does not match the TARA goal-set version. Action: flag the mismatch and hold the policy until it is reconciled. Checked IEC 62443-3-3 SR 3.4 (detect and report an integrity violation of configuration data), IEC 62443-2-1 CM 1.4 (change control for an IACS configuration change), ISO/SAE 21434 Clause 9 (goals as a work product, with traceability among work products), and ISO/IEC 27001 A.8.9 and A.8.32 (configuration shall be established and reviewed; changes shall pass change management) [1,7-9]. Each governs the file, the change, or the work product. None requires the deployed rule set to carry the TARA version, or a hold when the two differ. No match. The flag is retained.

NAC-R-08d has no match. Trigger: a certificate authenticates and its role attribute matches no defined tier. Action: place the session in quarantine and raise an alert. Checked IEC 62443-3-3 SR 1.1 and SR 1.2 (identify and authenticate), SR 1.11 (unsuccessful login attempts), SR 2.1 (enforce assigned authorizations; reject an operation the permissions do not cover), SR 2.1 RE(2) (define the permission-to-role mapping), and SR 5.2 RE(1) (deny network traffic by default, allow by exception); IEC 62443-2-1 ORG 1.3 (security roles and responsibilities); ISO/SAE 21434 Clauses 7 and 9; and ISO/IEC 27001 A.5.15, A.5.18, A.8.2, and A.8.5 [1,7-9]. SR 1.11 is failed authentication, which is NAC-R-07c. SR 5.2 RE(1) is traffic at a zone boundary, not a role attribute. SR 2.1 rejects an unauthorized operation; it does not quarantine a successfully authenticated principal whose role is undefined, and it does not require an alert. No match. The flag is retained.

Table 2 applies the overlay to the remaining rules. The substrate column repeats the binding already shown in Table 4, including the [38] basis of 07c and the [37] basis of 08a, so the two tables can be read together. Rules that share a trigger class and an overlay binding are grouped.

**Table 2.** Automotive overlay.

| Rules | Substrate | ISO/SAE 21434 | ISO/IEC 27001 |
|---|---|---|---|
| 01a, 05a, 05b | SR 1.1, 1.2, 1.9 [7] | Cl. 9 goals; Cl. 7 supplier interface [1] | A.5.16 identity; A.8.5 authentication [9] |
| 07c | [38] Sec. 5 | Cl. 9 goals; Cl. 7 supplier interface [1] | A.5.16 identity; A.8.5 authentication [9] |
| 08a | [37] §3.31 | Cl. 7 tiers; Cl. 9 goals [1] | A.5.15 access control; A.8.22 segregation [9] |
| 08f, 08b, 08c, 12a | SR 2.1; SR 5.1, 5.2 [7]; [38] for 12a | Cl. 7 tiers; Cl. 9 goals [1] | A.5.15 access control; A.8.22 segregation [9] |
| 08e | SR 2.8 [7] | Cl. 9 goals [1] | A.8.15 logging [9] |
| FC-03/04, FC-05/06/15/16, DEFAULT | SR 2.1 [7]; [39] for writes, [38] for DEFAULT | Cl. 9 goals CG-17, CG-18 [1] | A.8.3 information access restriction [9] |
| 02a | NIST CM-5 [39] | No 21434 NAC duty | A.8.2 privileged access [9] |
| 02b | CM 1.4 [8] | No 21434 NAC duty | A.8.32 change management [9] |
| 02c | SR 3.4 [7] | No 21434 NAC duty | A.8.9 configuration management [9] |
| 02d | SR 1.5; CM 1.4 [7,8] | No 21434 NAC duty | A.5.17 authentication information [9] |
| 02e † | None | None | None |
| 08d † | None (cf. [38]) | None | None |

† Flag of the Rules Without a Standard Equivalent subsection; no cited requirement covers the trigger and the action together.

Two readings of Table 2 matter. First, the 21434 column is an overlay, not a derivation. Clause 9 is why the goal exists. It is not why the rule has the trigger it has. Second, 27001 never supplies the system requirement. A.5.15 says access rules shall be established. It does not say which VLAN a tier receives, or which Modbus function an admitted session may invoke. That is the 62443 column.

**Energy Overlay**

The energy overlay replaces 21434. It does not replace the substrate. The example assumes medium-impact BES Cyber Systems with external routable connectivity under CIP-002-5.1a, which brings CIP-005-7 into force [40,41]. The bindings are to CIP-005-7 as cited; a later edition would need its Part numbers rechecked. ISO/IEC 27019:2024 supplies the energy-sector control guidance for process control, telecontrol, and remote maintenance [42]. If the site were low impact, CIP-005 would not apply as written and this column would change; the 62443 bindings would not.

NAC-R-E-06 has no match. Trigger: a vendor credential authenticates and its role is neither the PCS vendor nor the BMS vendor. Action: quarantine and raise an alert. This is the energy form of 08d, so the automotive check carries over: no match in IEC 62443-3-3, IEC 62443-2-1, or ISO/IEC 27001 [7-9]. Added for this rule, CIP-005-7 Part 1.3 requires inbound and outbound access permissions and denies all other access by default; Parts 2.4 and 2.5 require a method to determine and to disable active vendor remote access; Parts 3.1 and 3.2 require a method to determine and to terminate authenticated vendor-initiated connections to EACMS and PACS [41]. ISO/IEC 27019 brings remote maintenance systems into scope [42]. Part 1.3 is a network permission at the electronic access point, not a role attribute on a valid credential. Parts 2.4, 2.5, 3.1, and 3.2 act on a session that is already active. None quarantines an authenticated vendor whose role matches no defined equipment role, and none requires the alert. No match. The flag is retained. Table 6 therefore records no CIP-005 binding for E-06.

Vendor-session control under CIP-005 is a support claim, not a match for this trigger and action.

E-05 is the case where the overlay is easy to under-specify. A legacy BMS module or meter that is an applicable Cyber Asset on a routable protocol must reside in a defined electronic security perimeter (Part 1.1). External routable connectivity to it must pass an electronic access point (Part 1.2). The access point must deny all other access by default (Part 1.3) [41]. Those three parts cover the trigger, a path to the legacy asset, and the action, direct routes denied. The protocol proxy itself is not a CIP-005 requirement. It remains the 62443 compensating countermeasure of the Derived Rules subsection [7]. Table 3 and Table 6 therefore bind E-05 to CIP-005-7 Parts 1.1, 1.2, and 1.3 [41], and to ISO/IEC 27019 process control [42]. With that binding, E-01 to E-05 each have a 62443-3-3 requirement and a CIP-005 obligation, and none is flagged on its own.

Table 3 gives the overlay for E-01 to E-06. The substrate column repeats the binding in Table 6.

**Table 3.** Energy overlay, medium-impact assumption [40].

| Rules | 62443-3-3 | NERC CIP-005-7 | ISO/IEC 27019 |
|---|---|---|---|
| E-01 | SR 2.1 | Part 1.3, access permissions, deny by default | Process control; remote maintenance |
| E-02 | SR 1.1, 2.1, 5.2 | Parts 2.1 and 2.3, intermediate system and MFA; 2.4 and 2.5, vendor session | Remote maintenance |
| E-03 | SR 2.1, 5.2 | Part 1.3, access permissions | Process control |
| E-04 | SR 2.1 | Part 1.3, access permissions | Telecontrol |
| E-05 | SR 5.2 | Parts 1.1, 1.2, 1.3: asset in the ESP; external routable connectivity through an EAP; deny by default | Process control |
| E-06 † | None (cf. [38]) | None (see flag check) | None |

† Flag of the Rules Without a Standard Equivalent subsection. The 27019 column names the scope element the rule falls under, not a control identifier. The 27019 controls were not checked.

The CIP column is an obligation, not a copy of the rule. Part 1.3 applies at an electronic access point (EAP) and requires documented access permissions and a default deny there. The example site treats the site controller proxy, the zone boundary, and the grid gateway proxy as the EAPs of its electronic security perimeter. It does not name a control-room role, a DNP3 function, or an update role. Those selections come from the TARA goal and are bound on the substrate side by SR 2.1. The same limit applies to 27019. Its scope includes process control, telecontrol, and remote maintenance, which is why those words appear in Table 3. It is sector guidance for the control set. It is not a TARA method and not a rule generator. The column names the scope element rather than a control identifier.

The supporting rules are not renumbered (see Derived Rules). Their bindings are those of Table 2, with one addition: 02b and 02d also sit under CIP-010-4 R1 Part 1.2, which requires changes that deviate from a baseline configuration to be authorized and documented [43]. Part 1.2 applies to medium-impact BES Cyber Systems and their associated EACMS. CIP lists authentication servers among its examples of EACMS [41,43]. The example site treats the RADIUS server and the policy host as EACMS associated with its BES Cyber Systems. NAC-R-02e is reused as written and remains flagged. Nothing in CIP-005-7 or CIP-010-4 ties a deployed policy version to a risk-assessment version [41,43].

### What the Flags Decide

Three rules keep the flag: 02e, 08d, and E-06. E-06 is an instance of the 08d type, not a third type. The flags are not a claim of novelty against practice. They record that a reader cannot reach the rule by searching the cited standards for a requirement that states the trigger and the action together. The flags are retained in the related work, the rule tables, and the discussion that follows. CIP-005 vendor-session control and SR 2.1 are not treated as a match for 08d or E-06.

## Worked Example 1: Automotive OEM Multi-Tier Supplier Network

### Scenario and Item

The first example is an automotive OEM plant network with IT and OT zones. Suppliers in three tiers need remote access for different purposes. Legacy PLCs in the OT zone accept Modbus TCP commands without authenticating the sender. The item is the OEM-controlled network only: the boundary firewall, the RADIUS-based NAC, the PKI, the IT/OT segmentation rules, the HMIs, and the legacy PLCs. Supplier networks sit outside the item because the OEM cannot control them; supplier sessions are treated as external principals arriving at the boundary.

ISO/SAE 21434 is written for road vehicles [1]. Here its TARA method is applied to a non-vehicle item, the plant network that builds the vehicle, and its provisions for distributed cybersecurity activities frame the supplier tiers. The full TARA is in the appendix (Tables A1 to A4).

Enforcement points are named by role rather than product. In the reference implementation they are FreeRADIUS for the RADIUS server, OPNsense for the zone boundary, and a Python Modbus proxy for the protocol layer. The companion paper evaluates that implementation.

### Applied TARA Outputs

The TARA defines seven functions (F1–F7, Table A1) and eleven in-scope assets (A1–A11, Table A2). STRIDE applied per function–asset pair gives 37 threat entries (Table A3), each recording whether confidentiality, integrity, or availability is affected, which produce 24 cybersecurity goals (CG-01 to CG-24, Table A4). Impact is rated in the four ISO/SAE 21434 categories, safety, financial, operational, and privacy, on the standard's scale of severe, major, moderate, and negligible [1]. The overall impact of a threat is its highest category rating. Attack feasibility is not rated, because it needs site-specific operational data a general method cannot assume; every threat is treated as requiring treatment on impact alone, which is the conservative choice. The Limitations section records this as a limitation.

Eleven goals are recorded as future work and stay in the goal table (Table A4) as unmapped (Step 1). The other thirteen are in scope. Eleven of these yield rules; two are left unmapped with a reason. Three derivations are walked through below, chosen to cover both rule categories and both flagged rules. The remaining in-scope goals are treated briefly in the Remaining In-Scope Goals subsection.

**Derivation 1: Identity and Segmentation (CG-08)**

CG-08 states that authenticated entities shall be assigned to least-privilege VLANs based on verified identity and role. It comes from two Elevation of Privilege entries, F1/A9 and F4/A6 (Table A3, entries 15 and 16). In the second, a VLAN misassignment places an attacker in the OT VLAN with unrestricted reach to the PLCs.

CG-06 asks for the same decision from the supplier side: Tier-1, Tier-2, and Tier-3 credentials map to least-privilege zones. The two goals are derived together. What is protected: the segmentation rules at the IT/OT boundary (A6) and the enforcement decision itself (A9). Who is the subject: each supplier tier, and any principal whose role is not defined. How: the RADIUS server returns a VLAN with the access decision, and the zone boundary drops traffic that leaves the assigned zone.

Six rules follow (Table 4). NAC-R-08a places a Tier-1 supplier in VLAN 10, with access to the designated IT/OT conduit and no direct OT routing, using the VLAN tunnel attributes of RFC 3580 [37]. NAC-R-08f places a Tier-2 supplier in VLAN 20, with access to designated supplier data-exchange endpoints and no route to the IT/OT conduit or the OT segment. NAC-R-08b places a Tier-3 supplier in VLAN 30, limited to read-only data endpoints. Rules 08f and 08b are bound to SR 2.1 [7]. Only Tier-1 reaches the conduit; trust narrows from Tier-1 to Tier-3. VLAN numbers are illustrative. NAC-R-08c blocks and logs any attempt to reach a VLAN outside the assigned tier, under SR 5.2 [7]. NAC-R-08d is the default branch for the role attribute (see Rules Without a Standard Equivalent). NAC-R-08e logs every assignment, under SR 2.8 [7].

Consolidation changes the count. CG-07, which requires NAC decisions to be derived from documented TARA goals, produced three rules identical to 08a, 08b, and 08e. They are merged, and each merged rule now traces to both goals. CG-07 had no Tier-2 rule, so at this stage 08f traces to CG-06 and CG-08 only; CG-17 and CG-20 join its trace later, as Table 4 shows. The merged audit rule also records the policy version, which CG-07 asked for. NAC-R-07c is not merged into 08d. Both quarantine, but 07c fires on a failed authentication and 08d on a successful one with an unknown role. The action is the same; the trigger is not.

**Derivation 2: Two Enforcement Points (CG-17, CG-18)**

CG-17 states that all command sources to OT devices shall be authenticated before commands are accepted, through a proxy or an equivalent control. CG-18 states that OT device commands shall be validated against an authorized command set before execution, and that unauthorized writes shall be blocked. Both come from entries on F5/A1, the legacy PLC (Table A3, entries 27 to 29): a spoofed engineering workstation sending Modbus commands (Spoofing), a write that alters setpoints (Tampering), and an escalation from read to write access (Elevation of Privilege).

What is protected: the PLC and the process it controls. Who is the subject: any session that reaches the PLC's Modbus port, including a supplier session admitted legitimately. How: the PLC cannot authenticate the sender, so FR 1 fails at the endpoint and placement moves to the conduit (Step 3). Two decisions are needed. The first is who reaches the conduit at all. The second is which Modbus functions an admitted session may use.

The first decision needs no new rule. The identity rules of Derivation 1, with 07c, already keep any unauthenticated or unclassified session out of the OT conduit, and only Tier-1 is assigned to it. Consolidation adds CG-17 to their trace.

The second decision is written at a protocol proxy on the supplier conduit, in front of the PLC (Figure 2). Reads of holding and input registers, function codes 3 and 4, are permitted and logged. Writes, function codes 5, 6, 15, and 16, are denied. The proxy returns a Modbus exception response, so the client fails at once rather than waiting on a timeout. It logs the source connection, the function code, the rule that fired, and the rule's TARA reference. The proxy sees only the connection the overlay hands it, so the session's identity is recovered by correlating that connection with the overlay's own logs. A function code not in the policy is denied by a default rule, NAC-R-DEFAULT, with the same exception response. A frame the proxy cannot parse, or a PLC it cannot reach, closes the connection rather than passing traffic. These rules discharge use control, SR 2.1 [7], and least privilege as stated in [39], within the availability constraints of OT networks [38].

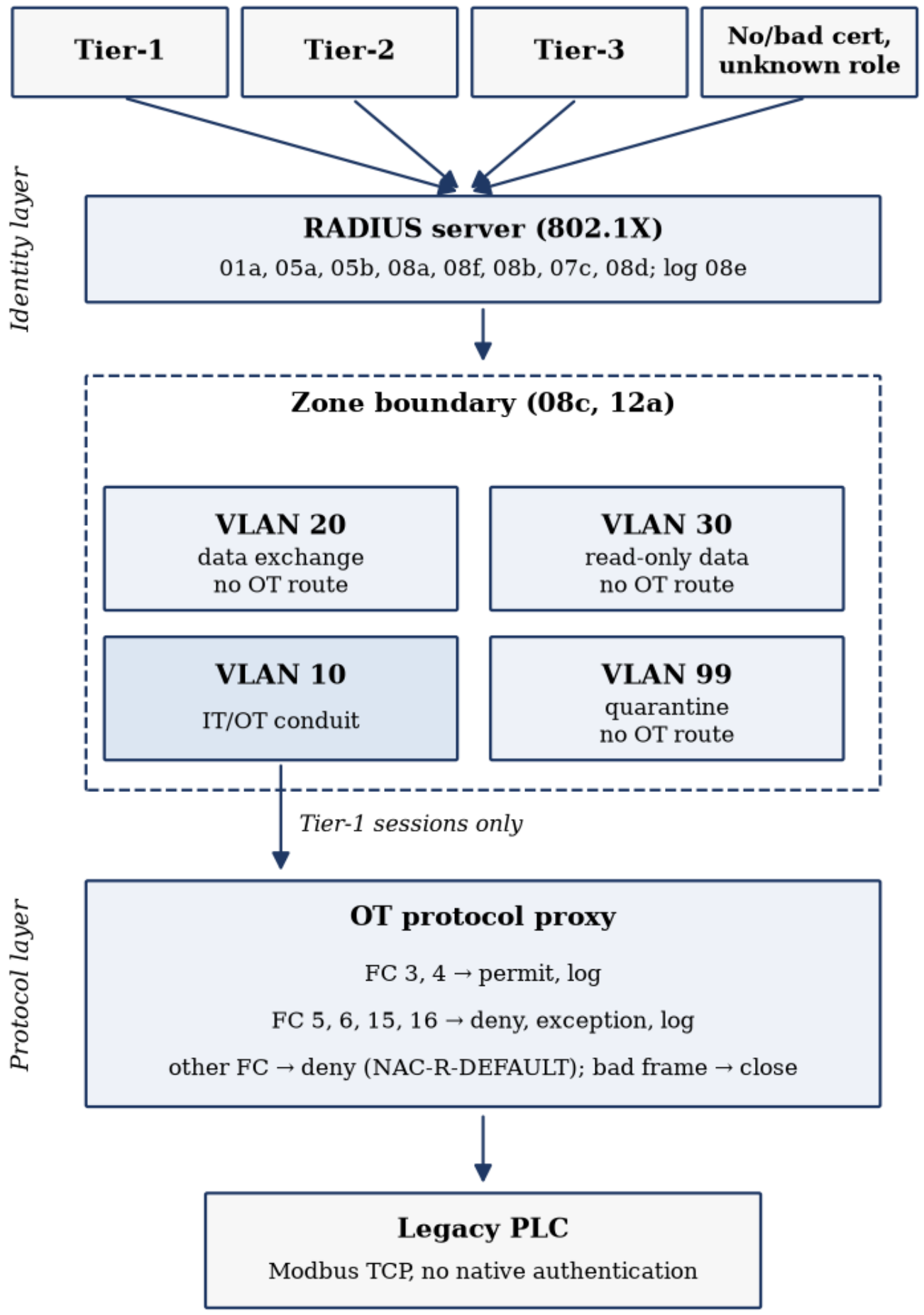

**Figure 2.** Two enforcement points for CG-17 and CG-18. The identity layer decides who reaches the OT conduit; the protocol proxy decides which Modbus functions an admitted session may invoke.

Sessions inside the OT zone are out of scope here: inside means a session that starts on a host in the OT zone, such as an engineering workstation or HMI, and reaches the PLC without crossing the supplier conduit. The proxy does not see these sessions, so CG-17 is only partly discharged in this example. Restricting them falls under CG-22, OT lateral movement, which is recorded as future work.

The CISA advisory shows why the second decision is not redundant. The activity of concern included read and write operations on data blocks and program logic [14]. A supplier admitted for monitoring holds the same session whether it reads or writes. CG-17 and CG-18 together produce rules at two enforcement points, which is the property claimed in the From Risk Analysis to Enforced Rules subsection.

### Derivation 3: Governing the Rule Set (CG-02)

CG-02 states that NAC policy rules and credential stores shall be integrity-protected and change-controlled. It comes from two Tampering entries, F1/A5 and F4/A10 (Table A3, entries 2 and 3). In the second, an insider changes the RADIUS or firewall policy so that an attacker is assigned to the wrong VLAN with elevated access.

What is protected: the policy and the credential store that every other rule depends on. Who is the subject: administrators, and anyone with access to the policy host. How: these are supporting rules (see Rule Taxonomy); they make no session decision.

NAC-R-02a restricts write access to policy files to authorized administrators [39]. NAC-R-02b allows a change only through documented change control and logs it with the before and after state [8]. NAC-R-02c compares the policy file hash with a known-good baseline and alerts on a mismatch [7]. NAC-R-02d requires dual authorization for changes to the credential store [7,8]. NAC-R-02e binds the policy to the TARA version (see Rules Without a Standard Equivalent). The change-control rule from CG-07, 07e, duplicates 02b and is merged.

NAC-R-02e closes the loop in Figure 1. The rules in Derivations 1 and 2 are only as good as their deployed state, and NAC-R-02e makes that state answer to the current TARA rather than to an earlier one.

### Remaining In-Scope Goals

Seven in-scope goals were not walked through (all goals are listed in Table A4). Five yield rules, three of them mostly by consolidation. Two are left unmapped, each with a reason (Step 6).

*CG-01* requires devices and users to be authenticated before network access is granted. NAC-R-01a requires 802.1X authentication on every access port and supplier entry point. No traffic other than the authentication exchange passes until it completes, and MAC-based bypass is disabled. This is the admission decision the tier rules assume, bound to SR 1.1 and SR 1.2 [7] and operated under [37]. A failed or absent authentication falls to 07c.

*CG-05* requires supplier identity claims to be verified with PKI certificates. NAC-R-05a accepts a certificate only if its chain validates to the OEM-trusted supplier CA and it is within its validity period. NAC-R-05b checks revocation status on each authentication. A revoked certificate, or one whose status cannot be confirmed, is rejected and logged. Both bind to SR 1.9, strength of public key authentication, which covers chain and revocation checks [7]. Rejections fall to 07c.

*CG-11* requires segmentation rules to be version-controlled and mapped to standards. It adds no rule. Change control with before-and-after records (02b) and policy-version binding (02e) already apply to the boundary configuration, and the standards mapping is the crosswalk in the Unified Standards Compliance Mapping section. Both rules now trace to CG-11.

*CG-12* requires the OT segment to be unreachable from IT or supplier zones without NAC enforcement. NAC-R-12a denies all traffic into the OT segment by default and permits only flows listed explicitly. Supplier flows are permitted only from the Tier-1 zone, through the conduit and the protocol proxy. It binds to SR 5.1 and SR 5.2 [7] and to the deny-by-default architecture of [38]. Rules 08c and 08d also trace to CG-12.

*CG-20* requires PLC register reads to be restricted to authorized systems. It adds no rule. Only Tier-1 reaches the conduit (08a), the other tiers are kept off it (08f, 08b), and reads reach the PLC only as function codes 3 and 4 through the proxy, behind 12a. Those rules now trace to CG-20.

*CG-14* requires every supplier connection to be authenticated with mutual TLS or an equivalent mechanism. That is a property of the transport carrying the supplier session to the OEM boundary, not a decision made at an enforcement point inside the item. It is left unmapped here. The companion paper implements it as a mutually authenticated zero-trust overlay and tests it.

*CG-24* requires OEM manufacturing IP to be unreachable from supplier sessions without explicit authorization. Zone assignment limits which hosts a supplier can reach (08a, 08f, 08b), but which records on those hosts a supplier may read is decided by the application or data store. That is outside NAC scope, so CG-24 is left unmapped with zone assignment noted as partial support.

**Result**

Table 4 lists the result. The sixteen rules derived for CG-02, CG-06, CG-07, and CG-08 consolidate to twelve. The protocol proxy adds seven: one per listed function code and a default deny for any other. The remaining goals add four more (01a, 05a, 05b, 12a). That gives 23 distinct rules covering eleven goals, with two in-scope goals unmapped for stated reasons. No conflicts arose during consolidation; every merge was an exact duplicate. Two rules carry the no-equivalent flag, 02e and 08d. Every rule in Table 4 has a backward trace to a threat entry, and every trigger class has a default branch. The bindings shown are the substrate bindings to IEC 62443 and NIST guidance; the Automotive Overlay subsection applies the ISO/SAE 21434 and ISO/IEC 27001 overlay to the same rules.

**Table 4.** Consolidated automotive rule set.

| Rule | CG | Trigger → action | Point | Basis |
|---|---|---|---|---|
| NAC-R-01a | 01 | Any port or entry point → 802.1X required; no MAC bypass | RADIUS | [7] SR 1.1, 1.2; [37] |
| NAC-R-05a | 05 | Supplier certificate → accept only if chain and validity check | RADIUS | [7] SR 1.9 |
| NAC-R-05b | 05 | Each authentication → revocation check; revoked or unknown → reject | RADIUS | [7] SR 1.9 |
| NAC-R-08a | 06, 07, 08, 17, 20 | Tier-1 certificate → VLAN 10; IT/OT conduit only | RADIUS | [37] §3.31 |
| NAC-R-08f | 06, 08, 17, 20 | Tier-2 certificate → VLAN 20; data-exchange endpoints; no conduit | RADIUS | [7] SR 2.1 |
| NAC-R-08b | 06, 07, 08, 17, 20 | Tier-3 certificate → VLAN 30; read-only endpoints | RADIUS | [7] SR 2.1 |
| NAC-R-07c | 01, 05, 07, 17 | No or invalid certificate → VLAN 99 | RADIUS | [38] Sec. 5 |
| NAC-R-08d † | 06, 08, 12, 17 | Valid certificate, unknown role → VLAN 99; alert | RADIUS | None (cf. [38]) |
| NAC-R-08c | 08, 12 | Cross-tier attempt → block; log | Zone boundary | [7] SR 5.2 |
| NAC-R-12a | 12, 20 | Any flow into OT → deny unless listed; suppliers via conduit and proxy only | Zone boundary | [7] SR 5.1, 5.2; [38] |
| NAC-R-08e | 07, 08 | Any assignment → log CN, role, VLAN, policy version | RADIUS; boundary | [7] SR 2.8 |
| NAC-R-FC-03/04 | 17, 18, 20 | FC 3, 4 read → permit; log | OT proxy | [7] SR 2.1 |
| NAC-R-FC-05/06/15/16 | 17, 18 | FC 5, 6, 15, 16 write → deny; exception; log | OT proxy | [7] SR 2.1; [39] |
| NAC-R-DEFAULT | 17, 18 | Any other FC → deny; exception; log | OT proxy | [7] SR 2.1; [38] |
| NAC-R-02a | 02 | Policy file write → administrators only | Policy host | [39] CM-5 |
| NAC-R-02b | 02, 07, 11 | Policy change → change control; log before/after | Policy host | [8] CM 1.4 |
| NAC-R-02c | 02 | Scheduled check → hash vs. baseline; alert | Policy host | [7] SR 3.4 |
| NAC-R-02d | 02 | Credential store change → dual authorization | RADIUS store | [7] SR 1.5; [8] |
| NAC-R-02e † | 02, 11 | Policy version ≠ TARA version → hold; flag | Policy host | None |

† No direct equivalent in the cited standards (see Rules Without a Standard Equivalent). FC: Modbus function code. Unmapped in-scope goals: CG-14, CG-24 (see Remaining In-Scope Goals).

## Worked Example 2: Utility-Scale Battery Energy Storage

### Scenario, Item, and Sources

The second example is a representative utility-scale battery energy storage site, assembled from public reviews and standards [32-36,40-43] and describing no real installation, operator, vendor, or product. Its parts are the ones those reviews describe. Battery racks are managed by a battery management system (BMS). A power conversion system (PCS) of inverters connects them to the grid. A site controller and energy management system (EMS) dispatches charge and discharge. A gateway at the grid interconnection exchanges telemetry and control with the utility's SCADA or distributed-energy management system [33]. Equipment vendors maintain the PCS and BMS remotely, in some designs through a cloud connection [34], and operations contractors reach the site from corporate IT.

The item is the site network the owner-operator controls. Vendor clouds, contractor networks, and the utility control center are outside it, as supplier networks were in the Worked Example 1 section. Vendors take the place of suppliers. Their roles are tied to the equipment they service, a PCS vendor and a BMS vendor, rather than to a supply-chain tier.

The example is scoped down from the Worked Example 1 section. There is no full TARA workbook. Six threat scenarios are drawn from the public sources, each produces one goal, and every goal is treated as requiring risk reduction on the same conservative basis as before.

### Overlay Assumptions

The Overlay Standards for the Energy Example subsection requires the impact category to be stated before any CIP identifier is attached to a rule. The example assumes the site's BES Cyber Systems are categorized medium impact under CIP-002 [40] and have external routable connectivity. That brings CIP-005 into force [41]: an electronic security perimeter whose access permissions deny by default, interactive remote access through an intermediate system with multi-factor authentication, and the ability to detect and disable active vendor remote sessions. ISO/IEC 27019 supplies the energy-sector controls for process control and remote maintenance [42]. If the site were categorized low impact, CIP-005 would not apply as written. The overlay column of Table 6 would change, and E-02's intermediate system and multi-factor authentication would become a site choice rather than an obligation; the other rules would not change.

### Threat Scenarios and Goals

Scenarios E1 to E5 follow threats described in the BESS literature [32-36]. E6 is the energy form of the unclassified-role case in the Rules Without a Standard Equivalent subsection. Table 5 pairs each scenario with its goal.

**Table 5.** BESS threat scenarios and goals.

| ID | Threat scenario | Cybersecurity goal |
|---|---|---|
| E1 | Dispatch or setpoint command reaches the EMS through a compromised vendor maintenance channel | Dispatch and setpoint commands are accepted only from authorized control roles |
| E2 | Contractor or vendor credentials used to pivot from corporate IT into the site network and across equipment zones | Vendor access is segmented by role and enters through a controlled intermediate point |
| E3 | Unauthorized firmware or protection-setting change pushed to the PCS or BMS | Firmware and settings change only through the authorized update role |
| E4 | Unauthorized control command, or false telemetry, at the grid interconnection gateway | Control commands at the interconnection originate only from the utility's authorized session |
| E5 | Legacy BMS module or meter that cannot hold a credential is reachable directly | Legacy assets are reachable only through a compensating enforcement point |
| E6 | Vendor credential with an undefined role is admitted to the site network | A vendor credential with an undefined role receives no site access |

Goal CG-E0n derives from scenario En and yields rule NAC-R-E-0n.

### Derived Rules

The procedure of the Methodology section runs unchanged. Table 6 lists the result, with the substrate binding to IEC 62443-3-3 and the overlay obligation side by side.

*Identity and segmentation.* NAC-R-E-02 admits vendor sessions only through the intermediate system with multi-factor authentication, then places each by role: the PCS vendor in the PCS zone and the BMS vendor in the BMS zone, with no route between them or to the EMS control functions. This is the automotive tier pattern (08a, 08f, 08b) with roles keyed to equipment instead of supply-chain tier. NAC-R-E-06 is its default branch. A vendor credential whose role is neither goes to quarantine and raises an alert, which makes E-06 the energy instance of 08d.

*Protocol and command.* NAC-R-E-01 and NAC-R-E-04 are the energy form of the Modbus filter. At the site controller, E-01 accepts dispatch and setpoint writes only from the control-room operator role; vendor roles may read. At the grid gateway, E-04 accepts the DNP3 control functions Select, Operate, and Direct Operate only from the utility's control session, permits reads to authorized readers, and denies any other function by default. The point is the one made in the Derivation 2 subsection: an admitted session is not yet an authorized command. E-04 does not address false telemetry. That is a data-integrity problem better met by detection [35], so that part of E4 stays open.

*Placement.* NAC-R-E-03 limits firmware and settings transfer to the PCS and BMS to the update-management role at the zone boundary. Checking the package signature is the device's job and is outside NAC scope. NAC-R-E-05 makes a legacy BMS module or meter reachable only through a protocol proxy on its conduit. That is the placement move used for the legacy PLC (Step 3), treated here as a compensating countermeasure [7].

The supporting rules carry over without change. Policy integrity, change control, and policy-version binding (02a–02e) and assignment logging (08e) apply at the site's policy host as written, so they are not renumbered.

**Table 6.** Derived BESS rules.

| Rule | Trigger → action | Point | 62443-3-3 | Overlay |
|---|---|---|---|---|
| NAC-R-E-01 | Dispatch or setpoint write → control-room role only; vendor roles read | Site controller proxy | SR 2.1 | CIP-005 perimeter permissions [41]; [42] |
| NAC-R-E-02 | Vendor session → intermediate system, MFA; zone by role; no cross-zone route | RADIUS; zone boundary | SR 1.1, 2.1, 5.2 | CIP-005 interactive remote access [41]; [42] |
| NAC-R-E-03 | Firmware or settings transfer to PCS, BMS → update role only | Zone boundary | SR 2.1, 5.2 | CIP-005 perimeter permissions [41]; [42] |
| NAC-R-E-04 | DNP3 Select, Operate, Direct Operate → utility session only; read → authorized; other → deny | Grid gateway proxy | SR 2.1 | CIP-005 perimeter permissions [41]; [42] |
| NAC-R-E-05 | Legacy BMS or meter → reachable only via proxy; direct routes denied | Zone boundary; proxy | SR 5.2 | CIP-005-7 Parts 1.1–1.3 [41]; [42] |
| NAC-R-E-06 † | Vendor credential, undefined role → quarantine; alert | RADIUS | None (cf. [38]) | None (see Energy Overlay) |

† No direct equivalent in the cited standards (see Rules Without a Standard Equivalent). Assumes medium-impact categorization under CIP-002 [40].

### Result

Six domain rules follow, one per goal, plus the supporting rules reused as written. No rule needed a new category. E-02, E-03, and E-06 are identity and segmentation rules, E-01 and E-04 are protocol rules, and E-05 combines a segmentation rule with a proxy behind it. Against the test of the Rules Without a Standard Equivalent subsection, E-01 to E-05 each map to a requirement in IEC 62443-3-3 and to an obligation in CIP-005, so none is flagged on its own. E-06 is flagged, but it is an instance of the rule type introduced by 08d, not a new one. Between the two domains, the overlay column and the role vocabulary changed. The procedure and the two decision layers did not.

## Discussion

### What Carried Over

The procedure ran in the Worked Example 2 section exactly as written in the Methodology section. The same six steps produced the energy rules, the same three checks closed them, and every energy rule fell into one of the two decision layers of the Rule Taxonomy subsection, with E-05 using both. No new rule category was needed.

Several automotive rules reappeared in the energy example in a different form. The tier rules (08a, 08f, 08b) became role-to-zone rules for vendors in E-02, and the default branch for an unknown role (08d) became E-06. The placement used for the legacy PLC, where enforcement moves to the conduit because the device cannot hold a credential, was reused for the legacy BMS module and meter in E-05. The supporting rules (02a–02e, 08e) did not change at all, which is to be expected, since they govern the rule set rather than any particular session.

The case for a separate protocol layer also held up. In both examples, a session that has been admitted is still not free to send any command it likes. The Modbus filter in front of the PLC and the DNP3 filter at the grid gateway (E-04) make the same kind of decision on different protocols.

### What Had to Change

The vocabulary of roles changed first. Automotive roles follow a supply-chain tier, whereas energy roles follow the equipment a vendor services. The protocol changed too. In Modbus the write function codes (5, 6, 15, and 16) can be denied as a group. In DNP3, E-04 works on the control functions Select, Operate, and Direct Operate, and E-01 at the site controller decides on the role behind a write rather than on the function, so the control-room operator can write while vendors can only read. The overlay was the third change: NERC CIP and ISO/IEC 27019 took the place of ISO/SAE 21434, and the IEC 62443-3-3 bindings stayed as they were (Table 6). One step had no counterpart in the automotive example. CIP obligations depend on how a system is categorized [40], so the energy example had to assume an impact category before it could attach any CIP identifier to a rule (see Overlay Assumptions).

Any overlay whose applicability depends on categorization rather than on the domain will need the same step. The energy example also exercised less of the procedure. Each scenario produced one goal and one rule, so consolidation in Step 5 never had anything to merge. The scenarios have no impact ratings and no full TARA behind them. What the transfer shows, then, is that the procedure and its rule categories can be run in a second domain. It says little about how they would behave on a TARA as large as the automotive one. As in the automotive case, the flag test of the Rules Without a Standard Equivalent subsection marked one energy rule, E-06, and that rule is an instance of the type 08d introduced, so no new flagged type appeared. The crosswalks in the Automotive Overlay and Energy Overlay subsections found no matching requirement for 02e, 08d, or E-06, so all three keep their flags.

**Beyond These Two Domains**

Two examples are not enough to claim that the method will work in a third domain, but they do make clear what a third domain would have to provide. It would need a risk assessment that produces goals with treatment decisions. It would need external parties whose roles can be carried in a credential, so that identity rules and a default branch can be written for them. For goals that concern actuation, it would also need a protocol whose functions an enforcement point can parse and sort into reads and control. When one of these is missing, the procedure records the fact instead of forcing a rule, either by marking the goal out of NAC scope or by leaving it unmapped with a reason. That is what happened with CG-14, CG-24, and the false-telemetry part of E4.

The advisory cited in the Introduction supports the choice of examples more than it supports their results. It names Critical Manufacturing and Energy among the most-targeted sectors [14], and the two examples sit in those sectors. The activity it describes, however, runs over S7comm rather than Modbus or DNP3, and a protocol rule for S7comm would need a proxy that can parse that protocol, which neither example attempts. Each new sector would also bring its own overlay, and that overlay's applicability would have to be stated before any rule is bound to it, as was done for CIP.

**Where Enforcement Is Tested**

This paper derives rules and does not deploy them, so whether they behave as written in a running network is a separate question. The companion paper takes up that question for the automotive example. It evaluates the reference implementation named in the Scenario and Item subsection, implements CG-14 as a mutually authenticated zero-trust overlay, and uses the protocol proxy of the Derivation 2 subsection, which filters Modbus function codes, logs the source connection, and relies on the overlay router's logs to identify the session. No result from that work is used or claimed here, and no implementation of the energy rules is reported.

## Limitations

Both worked examples are desk-based, and none of the derived rules is deployed or tested in this paper. The energy example is a representative site assembled from public sources rather than a real installation. It has no full TARA behind it, its scenarios carry no impact ratings (see Scenario, Item, and Sources), and its CIP bindings rest on the assumption that the site is categorized medium impact under CIP-002 [40] and has external routable connectivity. The derivation is manual, and each step depends on the analyst's judgment. Impact is rated in the automotive example only, and attack feasibility is rated in neither (see Applied TARA Outputs), so no risk value is computed and every goal is treated as needing risk reduction. In the automotive example, 11 of the 24 goals are deferred to future work and two in-scope goals, CG-14 and CG-24, remain unmapped (see Remaining In-Scope Goals and Table A4). Coverage also has gaps. The protocol rules decide on the Modbus function code alone and place no limit on which registers a permitted read can reach.

Because the proxy sees a connection rather than a principal, identifying the session depends on correlating it with the overlay router's logs. Sessions that start inside the OT zone never cross the proxy, so CG-17 is only partly discharged (see Derivation 2), and no rule addresses the false-telemetry part of E4 (see Derived Rules). Checking the deployed configuration against the rule table is left to tools such as [26]. Finally, the no-equivalent flags on 02e, 08d, and E-06 rest on the search of the cited standards in the Unified Standards Compliance Mapping section. That search covers the clauses named there and is not exhaustive, so a matching requirement elsewhere in those standards would remove a flag.

## Conclusions and Future Work

This paper has presented a procedure that takes the cybersecurity goals from a TARA and turns them into NAC rules at named enforcement points, with each rule traceable back to the threat records behind it. The procedure has six steps. A goal about who may reach an asset produces an identity rule, and a goal about what an admitted session may do to that asset produces a protocol rule as well. The rules are bound to IEC 62443-3-3 and ISO/IEC 27001 as a fixed substrate, and the domain regulation is applied on top as an overlay. In the automotive example, the procedure took 37 threat entries and 24 goals and produced 23 rules covering 11 of the 13 in-scope goals; the other two are unmapped for stated reasons. In the energy example, a representative battery energy storage site, it produced six rules under a NERC CIP and ISO/IEC 27019 overlay, with the substrate left unchanged. Two of the automotive rules and one of the energy rules have no equivalent in the cited standards.

Future work follows from the limitations in the Limitations section. The most direct next step is empirical: the companion paper tests the automotive enforcement points, and the energy rules still need a site or testbed of their own. The method should also be tried in further domains, each with its overlay's applicability settled first, and the 11 deferred automotive goals, among them OT lateral movement (CG-22), remain to be mapped. Finally, the derivation here is done by hand. Automated derivation of the kind already shown for CAN rules [6] could be extended to two enforcement points and a standards overlay, and rating attack feasibility and checking deployed configurations against the rule table [26] would close two further gaps.

## Definitions/Abbreviations

| | |
|---|---|
| ACL | Access control list |
| AI | Artificial intelligence |
| BES | Bulk electric system |
| BESS | Battery energy storage system |
| BMS | Battery management system |
| CA | Certificate authority |
| CAN | Controller area network |
| CG | Cybersecurity goal |
| CIP | Critical Infrastructure Protection (NERC standards) |
| CISA | Cybersecurity and Infrastructure Security Agency |
| CN | Common name (certificate field) |
| CPU | Central processing unit |
| CSF | Cybersecurity Framework (NIST) |
| DNP3 | Distributed Network Protocol 3 |
| DOE | United States Department of Energy |
| DoS | Denial of service |
| EACMS | Electronic access control or monitoring systems |
| EAP | Electronic access point |
| EMS | Energy management system |
| EPA | United States Environmental Protection Agency |
| ESP | Electronic security perimeter |
| FBI | Federal Bureau of Investigation |
| FC | Modbus function code |
| FR | Foundational requirement (IEC 62443) |
| HMI | Human-machine interface |
| IACS | Industrial automation and control system |
| IP | Intellectual property; Internet Protocol (IP address) |
| ISMS | Information security management system |
| IT | Information technology |
| LAN | Local area network |
| LLM | Large language model |
| MAC | Media access control |
| MFA | Multi-factor authentication |
| NAC | Network access control |
| NERC | North American Electric Reliability Corporation |
| NIST | National Institute of Standards and Technology |
| NSA | National Security Agency |
| OEM | Original equipment manufacturer |
| OT | Operational technology |
| PACS | Physical access control systems |
| PCS | Power conversion system |
| PKI | Public key infrastructure |
| PLC | Programmable logic controller |
| RADIUS | Remote Authentication Dial-In User Service |
| SCADA | Supervisory control and data acquisition |
| SIEM | Security information and event management |
| SR | System requirement (IEC 62443-3-3) |
| STRIDE | Spoofing, Tampering, Repudiation, Information disclosure, Denial of service, Elevation of privilege |
| TARA | Threat analysis and risk assessment |
| TCP | Transmission Control Protocol |
| TLS | Transport Layer Security |
| UNECE | United Nations Economic Commission for Europe |
| VLAN | Virtual local area network |
| VPN | Virtual private network |
| WAN | Wide area network |
| ZTNA | Zero trust network access |

## Declaration of AI-Assisted Technologies

The authors disclose that Claude (Anthropic) was used for reviewing manuscript text and figures, for checking standards identifiers and citations against the published sources, and for identifying inconsistencies in drafts. The authors have reviewed and edited all content as needed and take full responsibility for the scientific integrity and authenticity of this article.

## Appendix

Tables A1 to A4 give the full automotive TARA that the worked example summarizes: functions, assets, the 37 threat entries in workbook order, and the 24 cybersecurity goals with their disposition. Table A5 positions this paper against the prior work reviewed in the Related Work section.

**Table A1.** Functions in scope (F1 to F7).

| ID | Function | Description | Primary zone |
|---|---|---|---|
| F1 | Network Access Control | Authenticate and authorize devices/users attempting to connect to the OEM network via 802.1X/RADIUS | IT Zone |
| F2 | IT/OT Boundary Enforcement | Segment and control traffic flows between IT and OT network zones via firewall rules and routing policy | IT/OT |
| F3 | Supplier Connection Management | Receive, validate, and terminate incoming supplier network connections at the OEM boundary | IT Zone |
| F4 | VLAN Assignment & Enforcement | Place authenticated entities into appropriate network segments based on identity and policy | IT Zone |
| F5 | OT Device Communication | Enable authorized commands and data to reach legacy PLCs and HMIs across the OT network segment | OT Zone |
| F6 | Audit & Logging | Record authentication events, access decisions, policy enforcement actions, and traffic flows | IT Zone |
| F7 | ZTNA Tunnel Termination | Terminate and validate incoming supplier tunnels from the zero-trust (ZTNA) overlay at the OEM-controlled endpoint | IT Zone |

Primary zone as recorded in the TARA workbook.

**Table A2.** Assets in scope (A1 to A11).

| ID | Asset | Zone | Description |
|---|---|---|---|
| A1 | Legacy PLC / OT Device | OT Zone | Manufacturing control device. Receives Modbus TCP commands. No native authentication capability. |
| A2 | HMI (Human Machine Interface) | OT Zone | Operator interface for monitoring and controlling OT devices. May expose engineering parameters. |
| A3 | OT Network Segment | OT Zone | VLAN and routing infrastructure isolating OT devices from IT and external zones. |
| A4 | Boundary Firewall | IT Zone | Gateway enforcing traffic policy between IT, OT, and supplier zones. Most critical single asset in scope. |
| A5 | RADIUS NAC Server | IT Zone | Authentication decision engine. Evaluates 802.1X credentials and issues VLAN/policy assignments. |
| A6 | IT/OT Boundary Segmentation Rules | IT Zone | Firewall rules, ACLs, and VLAN policies defining permitted traffic flows across zones. |
| A7 | Incoming Supplier Connection | Boundary | Network connection request arriving from external supplier. OEM controls only its own termination point. |
| A8 | Supplier Identity Claim | Boundary | Credential or certificate presented by supplier during connection attempt to the RADIUS server. |
| A9 | NAC Enforcement Decision | IT Zone | The VLAN assignment and access policy that the RADIUS server and boundary firewall apply to an authenticated entity. |
| A10 | NAC Policy & Enforcement Rules | IT Zone | Configured rules governing how authentication outcomes map to network access decisions. |
| A11 | OEM Manufacturing IP | IT Zone | Process parameters, production configurations, and proprietary manufacturing data transmitted across the network. |

Item: the IT/OT network boundary of an automotive original equipment manufacturer, limited to the OEM-controlled network. Supplier internal infrastructure, in-vehicle networks, and physical production machinery are out of scope.

**Table A3.** Threat entries (37), STRIDE applied per function–asset pair.

| No. | F/A | STRIDE | CIA | Damage scenario | Attack vector | Impact S / F / O / P | Overall | Goal |
|---|---|---|---|---|---|---|---|---|
| 1 | F1/A5 | Spoofing | I | Attacker spoofs a valid device identity to the RADIUS server, gaining unauthorized network access without valid credentials | Credential replay / MAC spoofing against 802.1X | Ng / Mj / Sv / Ng | Severe | CG-01 |
| 2 | F1/A5 | Tampering | I | Attacker tampers with RADIUS server policy rules or credential store, causing unauthorized entities to be granted access | Admin interface exploit or unauthorized config change | Ng / Mj / Sv / Ng | Severe | CG-02 |
| 3 | F4/A10 | Tampering | I | NAC policy rules are tampered with causing incorrect VLAN assignments that grant attackers elevated network access | Unauthorized modification of RADIUS server or boundary firewall policy | Ng / Mj / Sv / Ng | Severe | CG-02 |

| No. | F/A | STRIDE | CIA | Damage scenario | Attack vector | Impact S / F / O / P | Overall | Goal |
|---|---|---|---|---|---|---|---|---|
| 4 | F1/A5 | Repudiation | I | Authentication events are not logged, enabling an attacker to deny having accessed the network with no audit trail available | Absence of audit logging on authentication events | Ng / Mo / Mj / Ng | Major | CG-03 |
| 5 | F3/A8 | Repudiation | I | Supplier denies having performed a network action because no binding audit trail links their identity claim to specific network events | Absence of session-bound audit logging | Ng / Mo / Mj / Ng | Major | CG-03 |
| 6 | F6/A5 | Repudiation | I | Failed and successful authentication events are not centrally logged enabling credential stuffing and insider access to go undetected | Absence of SIEM integration on authentication events | Ng / Mo / Mj / Ng | Major | CG-03 |
| 7 | F1/A5 | Denial of service | A | Attacker floods the RADIUS server with authentication requests, causing NAC service outage and blocking all legitimate network access | RADIUS request flood / DoS | Ng / Mj / Sv / Ng | Severe | CG-04 |
| 8 | F1/A8 | Spoofing | I | Malicious actor forges or replays a supplier identity claim causing the RADIUS server to grant trusted supplier access to a threat actor | Credential replay or certificate forgery via ZTNA | Ng / Mj / Sv / Ng | Severe | CG-05 |
| 9 | F3/A8 | Spoofing | I | Supplier presents stolen or cloned credentials allowing an unauthorized entity to masquerade as a trusted Tier-1 supplier | Stolen certificate or credential presented to the RADIUS server | Ng / Mj / Sv / Ng | Severe | CG-05 |
| 10 | F7/A8 | Spoofing | I | Supplier presents expired or revoked PKI certificate that is accepted due to absence of certificate revocation checking at OEM boundary | Revoked/expired certificate presented to ZTNA endpoint | Ng / Mj / Sv / Ng | Severe | CG-05 |
| 11 | F1/A8 | Elevation of privilege | I | A Tier-3 supplier credential is accepted at Tier-1 trust level due to absence of supplier tier hierarchy in NAC policy, granting excessive privileges | Valid lower-tier credential presented at higher-trust boundary | Ng / Mj / Sv / Ng | Severe | CG-06 |
| 12 | F3/A9 | Elevation of privilege | I | Supplier is granted access beyond their authorized tier because NAC has no supplier trust hierarchy model, violating least privilege | Policy gap exploitation during supplier authentication | Ng / Mj / Sv / Ng | Severe | CG-06 |
| 13 | F7/A8 | Elevation of privilege | I | Compromised supplier account is used to gain higher network privileges than the supplier tier authorizes due to missing tier enforcement | Privilege escalation via compromised supplier credential | Ng / Mj / Sv / Ng | Severe | CG-06 |
| 14 | F1/A9 | Tampering | I | NAC enforcement decision is based on misconfigured rules, placing attacker in wrong VLAN with elevated privileges | Policy misconfiguration exploited during authentication | Ng / Mj / Sv / Ng | Severe | CG-07 |
| 15 | F1/A9 | Elevation of privilege | I | Authenticated entity is assigned to OT VLAN instead of supplier VLAN due to missing privilege separation in enforcement policy | Privilege escalation via VLAN misassignment | Ng / Mj / Sv / Ng | Severe | CG-08 |
| 16 | F4/A6 | Elevation of privilege | I | VLAN assignment misconfiguration places attacker directly in OT VLAN granting unrestricted PLC access | VLAN hopping or policy exploitation | Mj / Mj / Sv / Ng | Severe | CG-08 |
| 17 | F2/A4 | Tampering | I | Firewall rules are modified by attacker or insider, collapsing IT/OT segmentation and enabling direct access to OT devices | Unauthorized admin access to the boundary firewall management interface | Mj / Mj / Sv / Ng | Severe | CG-09 |
| 18 | F2/A4 | Denial of service | A | The boundary firewall is overwhelmed by traffic or targeted DoS attack, taking down the IT/OT boundary and creating a momentarily flat network | Network flood targeting the boundary firewall WAN or LAN interface | Mj / Mj / Sv / Ng | Severe | CG-10 |
| 19 | F2/A6 | Tampering | I | Segmentation rules are deleted or overwritten causing a flat network where IT and OT zones lose isolation | Config management interface access | Mj / Mj / Sv / Ng | Severe | CG-11 |
| 20 | F2/A3 | Elevation of privilege | I | Attacker gains access to OT network segment by exploiting boundary misconfiguration, enabling direct PLC communication | Boundary firewall bypass or VLAN hopping | Mj / Mj / Sv / Ng | Severe | CG-12 |
| 21 | F2/A3 | Information disclosure | C | OT network topology, PLC addresses, and device configurations are exposed to unauthorized IT or supplier zone entities | Network scan from IT or supplier VLAN reaching OT segment | Ng / Mo / Mj / Ng | Major | CG-13 |
| 22 | F3/A7 | Spoofing | I | Attacker impersonates a legitimate supplier connection endpoint, gaining an authenticated foothold inside the OEM network | IP/session spoofing of supplier connection at OEM boundary | Ng / Mj / Mj / Ng | Major | CG-14 |

| No. | F/A | STRIDE | CIA | Damage scenario | Attack vector | Impact S / F / O / P | Overall | Goal |
|---|---|---|---|---|---|---|---|---|
| 23 | F7/A7 | Spoofing | I | Attacker hijacks or replays a legitimate supplier ZTNA session token gaining trusted network access as an authenticated supplier | ZTNA session token replay or hijack | Ng / Mj / Mj / Ng | Major | CG-14 |
| 24 | F3/A7 | Denial of service | A | Attacker floods OEM supplier connection endpoint with connection requests, causing denial of legitimate supplier sessions | Connection flood against ZTNA or VPN endpoint | Ng / Mo / Mj / Ng | Major | CG-15 |
| 25 | F7/A7 | Denial of service | A | Attacker floods ZTNA endpoint with tunnel establishment requests causing denial of service to all legitimate supplier sessions | ZTNA handshake flood | Ng / Mo / Mj / Ng | Major | CG-15 |
| 26 | F4/A10 | Information disclosure | C | NAC enforcement rules are exposed to an attacker revealing exploitable policy gaps and VLAN structure | Unauthorized read access to NAC configuration | Ng / Mo / Mj / Ng | Major | CG-16 |
| 27 | F5/A1 | Spoofing | I | Attacker impersonates an authorized engineering workstation and sends Modbus TCP commands to PLC with no authentication challenge | Direct Modbus TCP command injection — no authentication on the PLC | Mj / Sv / Sv / Ng | Severe | CG-17 |
| 28 | F5/A1 | Elevation of privilege | I | Attacker escalates from read-only Modbus access to write access on PLC due to absence of command-level authorization | Modbus function code escalation (read → write) | Mj / Sv / Sv / Ng | Severe | CG-17 |
| 29 | F5/A1 | Tampering | I | Attacker sends malformed or malicious Modbus write commands altering PLC setpoints causing incorrect physical process execution | Modbus TCP write command injection to PLC register | Mj / Sv / Sv / Ng | Severe | CG-18 |
| 30 | F5/A1 | Denial of service | A | Attacker floods PLC with Modbus requests causing CPU overload and halting the production line | Modbus TCP request flood targeting PLC CPU | Mj / Sv / Sv / Ng | Severe | CG-19 |
| 31 | F5/A1 | Information disclosure | C | Attacker reads PLC register values and process parameters revealing proprietary manufacturing configurations | Modbus TCP read command to PLC registers | Ng / Mj / Mo / Mo | Major | CG-20 |
| 32 | F5/A2 | Spoofing | I | Attacker spoofs process data displayed on HMI causing operators to make incorrect manual interventions based on falsified readings | Man-in-the-middle on HMI communication channel | Mj / Mj / Sv / Ng | Severe | CG-21 |
| 33 | F5/A2 | Tampering | I | Attacker modifies HMI configuration or display logic causing persistent falsification of operator process views | Unauthorized HMI configuration change | Mj / Mj / Sv / Ng | Severe | CG-21 |
| 34 | F5/A3 | Elevation of privilege | I | Attacker with initial OT foothold moves laterally across OT segment reaching multiple PLCs and HMIs without further authentication | Lateral movement within flat OT segment | Mj / Sv / Sv / Ng | Severe | CG-22 |
| 35 | F6/A4 | Repudiation | I | Firewall rule changes and boundary traffic anomalies go unrecorded enabling attackers to deny actions with no forensic evidence | Absence of firewall change logging and traffic audit | Ng / Mj / Mj / Ng | Major | CG-23 |
| 36 | F6/A10 | Repudiation | I | NAC policy changes are not audited allowing insider threat actors to weaken enforcement rules without detection or accountability | Unauthorized NAC policy modification without audit trail | Ng / Mj / Mj / Ng | Major | CG-23 |
| 37 | F7/A11 | Information disclosure | C | Attacker exfiltrates OEM manufacturing process parameters and proprietary production configurations through a compromised supplier ZTNA tunnel | Data exfiltration over established malicious ZTNA session | Ng / Sv / Mj / Mo | Severe | CG-24 |

F/A is the function and asset (Tables A1 and A2). CIA is the property affected. Impact categories: S safety, F financial, O operational, P privacy; Ng negligible, Mo moderate, Mj major, Sv severe. Overall is the highest category rating. Attack feasibility is not rated (see Applied TARA Outputs).

**Table A4.** Cybersecurity goals (24) and their disposition in this paper.

| Goal | Cybersecurity goal statement | Disposition |
|---|---|---|
| CG-01 | Devices and users SHALL be authenticated using non-repudiable credentials before network access is granted | Rules derived |
| CG-02 | NAC policy rules and credential stores SHALL be integrity-protected and change-controlled | Rules derived |
| CG-03 | All authentication attempts SHALL be logged with non-repudiable timestamps for forensic traceability | Deferred to future work |
| CG-04 | NAC authentication service SHALL be resilient to request flooding and maintain availability under attack | Deferred to future work |
| CG-05 | Supplier identity claims SHALL be cryptographically verified using PKI certificates before access is granted | Rules derived |
| CG-06 | NAC policy SHALL enforce supplier trust tier hierarchy mapping Tier-1/2/3 credentials to least-privilege access zones | Rules derived |
| CG-07 | NAC enforcement decisions SHALL be derived from documented TARA cybersecurity goals mapped to explicit policy rules | Rules derived |
| CG-08 | Authenticated entities SHALL be assigned to least-privilege VLANs based on verified identity and role | Rules derived |
| CG-09 | Firewall rules SHALL be integrity-protected, and change-audited | Deferred to future work |

| Goal | Cybersecurity goal statement | Disposition |
|---|---|---|
| CG-10 | IT/OT boundary enforcement SHALL remain available under network stress conditions and have defined failsafe behavior | Deferred to future work |
| CG-11 | IT/OT segmentation rules SHALL be version-controlled and mapped to IEC 62443 and ISO/SAE 21434 clause requirements | Rules derived |
| CG-12 | Access to the OT network segment SHALL require explicit authorization and SHALL NOT be reachable from IT or supplier zones without NAC enforcement | Rules derived |
| CG-13 | OT network topology and device addressing SHALL NOT be discoverable from IT or supplier zones | Deferred to future work |
| CG-14 | All supplier connections SHALL be authenticated at the OEM boundary using mutual TLS or equivalent cryptographic mechanism | In scope, unmapped |
| CG-15 | Supplier connection endpoints SHALL implement rate limiting and connection throttling to prevent service exhaustion | Deferred to future work |
| CG-16 | NAC policy configurations SHALL be access-controlled and not readable by unauthorized entities | Deferred to future work |
| CG-17 | All command sources to OT devices SHALL be authenticated before commands are accepted — proxy or equivalent control required | Rules derived |
| CG-18 | OT device commands SHALL be validated against authorized command sets before execution — unauthorized writes SHALL be blocked | Rules derived |
| CG-19 | OT devices SHALL be protected from request flooding via rate limiting enforced at the network boundary | Deferred to future work |
| CG-20 | PLC register read access SHALL be restricted to authorized systems only via boundary enforcement | Rules derived |
| CG-21 | HMI communication channels SHALL be integrity-protected to prevent data falsification | Deferred to future work |
| CG-22 | Lateral movement within the OT segment SHALL be restricted via micro-segmentation or device-level access controls | Deferred to future work |
| CG-23 | All firewall rule changes and boundary traffic anomalies SHALL be logged to a tamper-evident audit store | Deferred to future work |
| CG-24 | OEM manufacturing IP and process configurations SHALL NOT be accessible from supplier network sessions without explicit authorization | In scope, unmapped |

Rules derived: 11 goals. In scope, unmapped: CG-14 and CG-24 (see Remaining In-Scope Goals). Deferred to future work: 11 goals, including CG-22.

**Table A5.** Positioning of this paper against the closest prior work, as described in the Related Work section.

| Work | Starts from | Produces | Standards mapped | Limit relative to this paper |
|---|---|---|---|---|
| Navarra et al. [6] | TARA artifacts | CAN firewall rules, derived by an automated, LLM-assisted workflow with explicit risk traceability | ISO/SAE 21434 | Every rule at one enforcement layer (the vehicle gateway); no supplier or organizational boundary; no mapping beyond ISO/SAE 21434; one domain |
| Wang et al. [5]; Dobaj et al. [2]; Wagner [17] | TARA results | Cybersecurity requirements [5]; design patterns [2]; cross-divisional risk metadata [17] | — | The TARA ends at requirements, patterns, or metadata; no enforcement rule |
| Luo et al. [18, 19] | Threat analysis | Gateway security mechanisms and firewall design | — | No documented derivation procedure |
| Kleberger et al. [20] | In-vehicle network | Security domains, partitioned algorithmically | — | Stops at the architecture |
| Djebbar and Sogbade [12] | Threats (no TARA) | List of controls to implement | ISO/IEC 27001, NIST CSF, IEC 62443-3-3 | No configured rules; no TARA input |
| Schachinger et al. (PURITY) [11] | — | Controls mapped onto Purdue-model levels | ISO/IEC 27001, IEC 62443, NIST SP 800-82, ISO/IEC 27033 | Not connected to a domain risk assessment; no ISO/SAE 21434 |
| Jaatun et al. [21]; Elmarkez et al. [22] | Zones and security levels; threat evaluation [22] | Principles and a process for connecting zones [21]; security functions per zone [22] | IEC 62443 | No per-identity rules for a conduit; functions identified per zone, not per rule at a named enforcement point |
| Vaudey et al. [23]; Kern et al. [24] | System design | Formal zone models: domain-specific language [23]; data-flow zone modeling [24] | IEC 62443 | Models the zones, not the rules enforced at them |
| Cheminod et al. [25, 26] | Policy (assumed to exist) | Policy description [25]; semi-automatic check that an implementation conforms to its specification [26] | — | Where the policy comes from is not addressed |
| Leander et al. [27]; Kern and Anderl [28]; Fink et al. (P4Ward) [29] | Policy (assumed to exist) | Enforcement architectures [27]; attribute-based enforcement at network and machine level [28]; device-specific behavioral enforcement at line rate [29] | — | The policy is assumed, not derived from a risk assessment |
| Costantino et al. [10]; Akkus and Annighoefer [13] | Two standards | Standard-to-standard comparison | UNECE R155 and ISO/SAE 21434 [10]; aviation and automotive standards [13] | No rules; comparison only |

| Work | Starts from | Produces | Standards mapped | Limit relative to this paper |
|---|---|---|---|---|
| Eschlberger et al. [30]; Rabe and Wollert [31] | Smart-grid analysis language [30]; software-update requirements [31] | Adaptation to automotive [30]; transfer to fire alarm systems [31] | ISO/SAE 21434 and UN R155 [30] | Transfer of an analysis language or of requirements, not of a rule-derivation procedure; [30] runs in the opposite direction |
| Battery energy storage security work [32–36] | — | Threat-modeling survey [32]; security review [33]; attack-surface demonstration [34]; detection and mitigation review [35]; resilient controller study [36] | — | No structured risk assessment carried through to access control rules |
| **This paper** | TARA cybersecurity goals | NAC rules at two enforcement points, each traceable to threat records and bound to a standard requirement or flagged where none exists | IEC 62443-3-3 and ISO/IEC 27001 (substrate); ISO/SAE 21434, or NERC CIP and ISO/IEC 27019 (overlay) | Derivation is manual; examples are desk-based; deployment is tested in the companion paper |

— Not addressed in the work as described in the Related Work section. Numbers in brackets refer to the reference list.